\documentclass[svgnames,x11names,sigconf]{acmart} 

\AtBeginDocument{%
  \providecommand\BibTeX{{%
    \normalfont B\kern-0.5em{\scshape i\kern-0.25em b}\kern-0.8em\TeX}}}

\copyrightyear{2024}
\acmYear{2024}
\setcopyright{acmlicensed}\acmConference[CHI '24]{Proceedings of the CHI Conference on Human Factors in Computing Systems}{May 11--16, 2024}{Honolulu, HI, USA}
\acmBooktitle{Proceedings of the CHI Conference on Human Factors in Computing Systems (CHI '24), May 11--16, 2024, Honolulu, HI, USA}
\acmDOI{10.1145/3613904.3641904}
\acmISBN{979-8-4007-0330-0/24/05}

\usepackage[]{xcolor}
\usepackage{soul} %
\usepackage{microtype}
\usepackage{framed}
\definecolor{shadecolor}{rgb}{0.9,0.9,0.9}
\usepackage{multirow}
\usepackage{makecell}
\usepackage{algorithm, algorithmic}
\usepackage{float} %
\usepackage{wrapfig} %
\usepackage{enumitem} %
\usepackage{xspace}

\makeatletter
 \def\SOUL@hlpreamble{%
 \setul{}{3.5ex}%
 \let\SOUL@stcolor\SOUL@hlcolor
 \SOUL@stpreamble
 }
\makeatother

\definecolor{TodoColor}{rgb}{1,0.7,0.6}

\newcounter{textexamplecounter}

\newcommand{\textexample}[4][\centering]{
    \vspace{2mm}
    \noindent
    
    \begin{minipage}{\linewidth}
    \begin{center}
    \colorbox{gray!10}{
    \begin{minipage}{0.95\linewidth}
    \refstepcounter{textexamplecounter}
    \vspace{2mm}
        \setlength{\parindent}{0cm}
        {
            \fontsize{0.95em}{0.95em}\selectfont
            #4
            
        }
        \vspace{0.5mm}        
    \vspace{1mm}
    \end{minipage}
    }
    \end{center}
    {
    #1
    {\bf \noindent Example \thetextexamplecounter: #3 \par}
    }
    \end{minipage}
}

\let\svthefootnote\thefootnote
\newcommand\blankfootnote[1]{%
  \let\thefootnote\relax\footnotetext{#1}%
  \let\thefootnote\svthefootnote%
}

\newcommand{\system}{\textit{RELIC}\xspace}
\newcommand{\view}[1]{\textit{#1}\xspace}

\definecolor{tblue}{RGB}{76,120,168}
\definecolor{torange}{RGB}{245,133,24}
\definecolor{revision}{RGB}{114,158,206}

\begin{document}

\title[RELIC: Investigating Large Language Model Responses using Self-Consistency]{RELIC: Investigating Large Language Model Responses using~Self-Consistency}

\author{Furui Cheng}
\affiliation{%
  \institution{ETH Zurich}
  \city{Zurich}
  \country{Switzerland}}
\email{furui.cheng@inf.ethz.ch}

\author{Vilém Zouhar}
\affiliation{%
  \institution{ETH Zurich}
  \city{Zurich}
  \country{Switzerland}}
\email{vzouhar@inf.ethz.ch}

\author{Simran Arora}
\affiliation{%
  \institution{Stanford University}
  \city{Stanford, California}
  \country{USA}}
\email{simarora@stanford.edu}

\author{Mrinmaya Sachan}
\affiliation{%
  \institution{ETH Zurich}
  \city{Zurich}
  \country{Switzerland}}
\email{msachan@inf.ethz.ch}

\author{Hendrik Strobelt}
\affiliation{%
  \institution{IBM Research}
  \city{Cambridge, Massachusetts}
  \country{USA}}
\email{hendrik.strobelt@ibm.com}

\author{Mennatallah El-Assady}
\affiliation{%
  \institution{ETH Zurich}
  \city{Zurich}
  \country{Switzerland}}
\email{menna.elassady@ai.ethz.ch}

\renewcommand{\shortauthors}{Cheng, Zouhar, Arora, Sachan, Strobelt, and El-Assady}

\begin{abstract}
Large Language Models (LLMs) are notorious for blending fact with fiction and generating non-factual content, known as hallucinations. To address this challenge, we propose an interactive system that helps users gain insight into the reliability of the generated text.
Our approach is based on the idea that the self-consistency of multiple samples generated by the same LLM relates to its confidence in individual claims in the generated texts.
Using this idea, we design \system, an interactive system that enables users to investigate and verify semantic-level variations in multiple long-form responses. This allows users to recognize potentially inaccurate information in the generated text and make necessary corrections. 
From a user study with ten participants, we demonstrate that our approach helps users better verify the reliability of the generated text.
We further summarize the design implications and lessons learned from this research for future studies of reliable human-LLM interactions.
\end{abstract}




\begin{teaserfigure}
  \includegraphics[width=\textwidth]{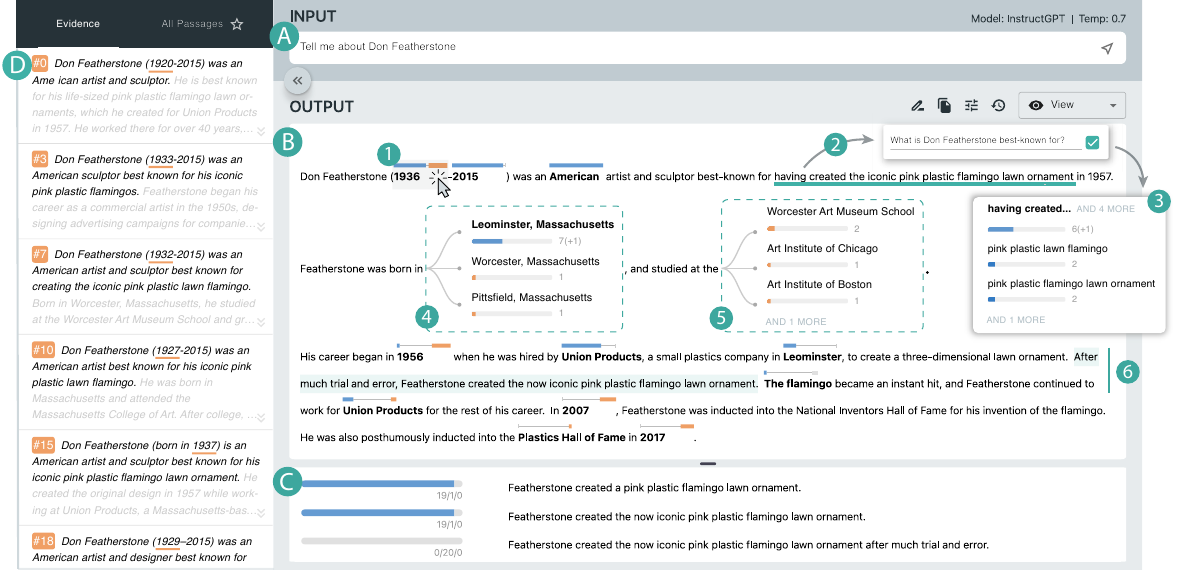}
  \caption{\system{} allows users to search for information from large language models (A), view the model's top response, and understand the variations between response samples to verify the correctness of the generated information (B).  For long-form generated text, users can inspect the consistency of each individual claim (C) and find \textcolor{torange}{contradicting} or \textcolor{tblue}{supporting} evidence from other samples (D). Steps (1-6) illustrate the user's verification process of InstructGPT's response regarding Don Featherstone.}
  \label{fig:teaser}
  \Description{The figure presents the interface design of RELIC and illustrates a use case in which a user verifies InstructGPT's response regarding a prompt, "Tell me about Don Featherstone." From this figure, we know that RELIC allows users to search for information from large language models like InstructGPT by inputting a prompt, viewing the model's top response, and understanding the variations between multiple response samples to verify the correctness of the generated information. For long-form generated text, users can inspect the variations on each piece of information, noted as atomic claims, and find contradicting or supporting evidence from other samples. For long-form generated text, users can inspect the variations on each piece of information, noted as atomic claims, and find contradicting (marked in orange) or supporting evidence (marked in blue) from other samples.}
\end{teaserfigure}

\maketitle

\section{Introduction}

Large Language Models (LLMs) have demonstrated impressive capabilities in generating highly fluent texts~\citep{fang2023chatgpt}.
LLM-built tools, such as chatbots, writing assistants, and search engines, are becoming increasingly popular and are used by hundreds of millions of people \citep{Choudhury2023InvestigatingTI}.
LLMs generally allow people to engage with the world of data and information using natural language.
Nevertheless, there is no guarantee that the model response---the generated information---is accurate.
On the contrary, LLMs are notorious for blending fact with fiction and generating non-factual content, known as hallucinations~\citep{ji2023survey}.
Despite the fact that the content is false or not adequately supported by sources, it is often presented in a convincing and fluent way. 
The seemingly realistic and convincing statements can mislead users and cause legal and ethical issues.
Preventing users from being misled by hallucinations has become one of the most urgent issues facing the community.

Previous human-AI interaction studies found that displaying the model's confidence level (usually represented with a score or score interval) is an efficient way to help users make correct decisions~\cite{zhang2020effect}. 
This idea can be extended to natural language generation (NLG) tasks. 
Intuitively, if the language model provides inconsistent or contradictory responses to the same prompt or a question, it indicates a lack of reliability.
In this case, users could disregard the generated content because it is unlikely to be true (see Example \ref{example:consistency}).

\vspace{-2mm}
\begin{figure}[htbp]
\textexample[\centering]{\linewidth}
{Inconsistency shows the lack of accuracy.}
{\label{example:consistency}
\it
\textbf{\#1}: Rodrigo is a footballer who plays for Man City.\\
\textbf{\#2}: Rodrigo is a footballer playing for Atlético Madrid.\\
\textbf{\#3}: Rodrigo is a Costa Rican actor and musician.
}
\end{figure}
\vspace{-2mm}

Although this scenario seems straightforward, it is difficult to measure and efficiently communicate consistency between text samples to users.
Unlike classes in classification problems, text generations are not mutually exclusive. 
Sentences that comprise different words can often have the same meaning, being semantically equivalent, and sentences with similar partial surface-level forms can have different meanings or might even be contradictory.
Therefore, the commonly used perplexity-based metrics computed using token-level log probabilities cannot be reliably used to estimate the model's confidence in the correctness of the generated text.
This problem becomes even more challenging in long-form NLG. %
It is hard to use a single score or score range to indicate its trustworthiness, as in traditional classification and regression problems.
In this study, we investigate a user-centric approach combining natural language processing (NLP) techniques and interactivity to help users comprehend the consistency between the multiple LLM outputs to verify and correct hallucinations.

We initialize the research (Section \ref{sec:formative_study}) with a formative study to understand the limitations of existing LLM interfaces (e.g., \href{https://platform.openai.com/playground}{\textcolor{black}{OpenAI Playground}}) in conveying LLM's confidence in the generations and the desired improvements from users. 
The study shows that users require a clear and concise visual summary to understand the model's confidence at the entire text level.
Additionally, \textit{interactive} and \textit{in-context} validations assist users in comprehending the model's confidence level more precisely. 
Finally, users expect to view concrete sentences that support or contradict each claim using an \textit{evidence-driven} approach in validation rather than just being told if it's right or wrong.
These findings help us form the requirements for enabling an efficient self-consistency understanding workflow.

We propose \system 
, an interactive system (Section \ref{sec:interface}) built upon a novel self-consistency-checking algorithm (Section \ref{sec:algorithm}) that helps LLM users identify and steer inaccurate information in the generated text by understanding the consistency between multiple responses.
In developing \system, we encountered three technical challenges: \textit{measuring semantic-level consistency}, \textit{linking consistency justifications with evidence}, and \textit{presenting this information to users effectively}. 
We created a computational pipeline (\autoref{fig:algo_main}) that breaks down the generation into atomic claims and uses an off-the-shelf natural language inference model to assess the support for each atomic claim from a list of additional samples as a measure of the model's self-consistency.

To allow users to interactively retrieve the words and sentences from additional samples that support and contradict certain claims (i.e., evidence), we propose a novel approach that integrates a question-answering-based pipeline with interaction techniques. 
For each atomic claim or user-selected piece of text (e.g., \textit{Rodrigo is a footballer.}), the system automatically generates a question in natural language (e.g., \textit{What is Rodrigo's profession?}) and uses this question to locate the answer tokens from all samples. 
We designed visualizations and interactions based on this technique that allows users to inspect the alternative options regarding each piece of information (e.g., if Rodrigo is not a footballer, what other professions could he pursue according to other samples) and interactively query supporting evidence for each option.  
Finally, we enable users to conduct a what-if analysis by editing the generated text and gaining new self-consistency judgments regarding the changes.

We conducted a user study (Section \ref{sec:user-study}) with ten participants from diverse backgrounds to evaluate the system's usability and usefulness.
In the user study, the users were asked to validate generations from InstructGPT using the system and provide quantitative and qualitative feedback regarding their usage experience. 
During the user study, we observed that users frequently navigated through samples to find evidence and edit the generated text for conducting what-if analyses.  
The results confirm the usability and usefulness of the system and illustrate the impact on users' workflow with LLMs, where users more actively interact with LLM generations,  playing roles beyond viewers but also of examiners and editors. 
To demonstrate the usage of the system, we present a case study (Section \ref{sec:case-study}) based on the experience of one participant.
Combining the results from the multi-method evaluations, we conclude that our approach enables a user-centered workflow for better verifying the generated text.

\vspace{3mm}
We highlight the three main contributions of our research:
\begin{itemize}[topsep=3mm,itemsep=4mm]
\item \textbf{A Formative Study}, illustrating the shortcomings of current LLM interfaces in indicating the reliability level of the generated text and helps us form requirements for designing a user-centered NLG verification system. 

\item \textbf{\system, an Interactive System} that helps users verify and steer NLGs from language models by investigating the factual consistency of multiple samples. 

\item \textbf{A User Evaluation} with ten participants demonstrating the effectiveness of the proposed approach and bringing insights for future human-LLM interaction studies.
\end{itemize}

\section{Related Work}\label{sec:related_work}
In this section, we review related research approaches, mainly focusing on the aspects of computational method designs for NLG factuality verification and interaction designs for human-LLM collaboration in NLG applications.

\subsection{Fact Verification and Correction in Natural Language Generations}\label{sec:fact_verification}
In recent years, researchers from the NLP field have proposed a variety of techniques to automatically assess and improve the faithfulness and factuality of LLM generations. 
These approaches have been summarized in recent survey papers~\cite{ji2023survey,wang2023survey,zhang2023hallucination}. 
In this section, we briefly introduce the common approaches and how they relate to our method.

\paragraph{Retrieval-Augmented Approaches}
A common approach leverages knowledge retrieved from external sources to verify and correct NLGs~\cite{thorne2018fever}.
For instance, \citet{petroni2022improving} proposes SIDE, a system that aims to improve the quality of citations for the claims in Wikipedia pages by retrieving and ranking alternate sources from a pre-curated document corpus.
Min et al. proposed FactScore~\citep{min2023factscore}, an evaluation framework to assess the factuality of long-form text by decomposing the generated text into atomic claims and verifying them individually using external knowledge sources. 
These methods benefit from external knowledge bases that can be audited and managed independently.
However, reliable knowledge resources might not always exist or be up-to-date.

\paragraph{Consistency-based Approaches}
A more general approach assesses the consistency among the generated text from the same context to gain more accurate outputs.
The concept of self-consistency is proposed by Wang et al.~\cite{wang2023selfconsistency} as a decoding strategy that samples multiple reasoning paths and selects the most common one to improve output quality.
Yao et al. extended the self-evaluation method and proposed Tree-of-Thought~\cite{yao2023tree}, a framework that optimizes LLMs' decision-making by sampling and evaluating the intermediate states in the reasoning process using consistency.

Consistency can also be utilized to evaluate and improve the factuality of LLM responses.
Kuhn et al.~\citep{kuhn2022semantic} proposed the \textit{Semantic Uncertainty} measurement, which uses an off-the-shelf natural language inference model to characterize the variations between multiple LLM responses given the same prompt. 
A similar approach named SelfCheckGPT~\citep{manakul2023selfcheckgpt} uses automatically generated multiple-option questions to evaluate the consistency of each sentence. 
A concurrent work, ChatProtect~\citep{muendler2023selfcontradictory}, removes the hallucinations by resolving the contradictions among multiple sampled responses.

There are other methods besides the ones mentioned above, such as fine-tuning the LLM with factuality preference rankings~\cite{tian2023fine} and using multiple LLM instances to debate their individual response to arrive at a superior answer~\cite{du2023improving}. 

Despite the vast efforts in automatically assessing and enhancing the faithfulness and factuality of LLM responses, there has been insufficient research into empowering end users' agencies in inspecting and comprehending the possible flaws in LLM responses.
From the end user's perspective, they hold the right to justify the reliability of LLM responses to decide which part to trust and thus should be equipped with useful tools to make the decisions.
Our study fills this gap through a human-centered approach.
Our method is built upon Semantic Uncertainty and FactScore, which helps users interactively comprehend and confirm factuality in the generated text through self-consistency.

\subsection{Human-LLM Interactions}
\label{sec:related_work_2}

Recent studies in human-LLM interaction have investigated various ways humans can collaborate with LLMs in accomplishing downstream tasks, like text summarization~\citep{cheng2022mapping}, creative writing~\citep{lee2022coauthor,ippolito2022creative,yao2021ai,yang2022ai}, and information seeking~\cite{jiang2023graphologue,suh2023sensecape}.

\paragraph{LLM's Roles in Collaboration}
Previous research has investigated how LLMs can aid in the process of writing~\citep{petridis2023anglekindling, ding2023mapping}. 
Yuan et al. proposed Workcraft~\citep{yuan2022wordcraft}, an interactive writing assistant tool powered by GPT-3. 
In Workcraft, the language model assists the writer by generating writing suggestions according to users' instructions, like \textit{Infilling}, \textit{Continuation}, and \textit{Elaboration}.
Using this tool as a probe, Ippolito et al.~\citep{ippolito2022creative} found that professional writers desire to use LLMs as brainstorming partners, research assistants, etc. 
Additionally, a study by Dang et al.~\citep{dank2022beyond} suggested that LLMs can be helpful commentators by providing continuous text summaries. 
Most existing work targets low-risk scenarios, such as creative writing, where users can easily decide to accept or reject suggestions from LLMs. 
Our study examines situations where LLMs cannot always be trusted as assistants, and users must verify the accuracy of the model's responses.

\paragraph{Controllability \& Transparency of LLMs}
Previous studies suggest controllability and transparency are two important pillars for building a responsible AI system~\citep{wang2019designing, DAS2023103219, fernandes2023bridging, liao2023ai}. 
Compared with conventional AI systems, LLMs enable more user-friendly interactions where users use natural language instructions (i.e., prompts) to gain model responses. 
However, to get desired outputs from LLMs, the prompt has to be well-calibrated, which is not easy for novice end users~\citep{jiang2022promptmaker,strobelt2022interactive,zamfirescu2023johnny}. 
To enhance the controllability of operating LLMs to achieve complex tasks, researchers propose a strategy called chain-of-thought~\citep{wei2022chain, wu2022ai, wu2022promptchainer} that uses LLMs to resolve elementary sub-tasks step-by-step and builds a pipeline by chaining the modules.
Despite the prompting process, another area of research aims to improve human controllability in navigating the complex LLM response~\citep{jiang2023graphologue,suh2023sensecape}.
For example, Jiang et al. proposed Graphologue~\citep{jiang2023graphologue}, an interactive system that allows users to explore model responses with interactive diagrams. 

Most of the existing studies approach the goal of transparency by providing explanations~\citep{lipton2018mythos}. 
Among these studies, a common approach is to allow users, mainly professional model developers, to look inside the model and understand the model's mechanism~\citep{sevastjanova2022visual, sevastjanova2022lmfingerprints, spinner2019explainer}, such as LSTMVis~\citep{strobelt2017lstmvis}, Seq2seq-Vis~\citep{strobelt2018s}, and LIT~\cite{tenney2020language}.
When the size and complexity of language models increase (e.g., with more attention layers), 
it's harder to understand the model's mechanism by opening the black box.  
An alternative approach is to probe language models with different prompts to understand the input-output associations~\citep{ribeiro2016should}. 
These model-agnostic approaches are mainly designed for classification tasks, like sentiment analyses. 
Extending these methods to explain long-form text generation remains an open question. 

In our study, we target a different problem where we aim to help users gain more authority and controllability for understanding and steering LLM \textit{responses} that are potentially hallucinatory. 
We propose that consistency between multiple samples provides transparency to LLM responses~\citep{bhatt2021uncertainty}.
To achieve this, we propose measurements and interactions to visually summarize the divergence between long-form model responses and enable users to efficiently navigate through them. 

\section {Informing the Design}
\label{sec:formative_study}

We conducted a formative study to learn how users search for and confirm factual information from LLM-generated text, as well as the shortcomings of current LLM interfaces (e.g., OpenAI Playground) in indicating the level of reliability in the generated text. 
From the study, we identify requirements for designing an efficient LLM interface to convey the reliability of the generated text.

\subsection{Participants and Study Procedure}
\label{sec:participants_study}

\paragraph{Participants.} 
We recruited five participants (aged 25 - 33) with various backgrounds, including one journalist, two PhD students studying Business, and two NLP researchers (not the authors). 
All participants reported using LLMs in their daily lives. 
The most common reasons for using them were to seek writing advice (four out of five participants), support text reading (four out of five), and search for information (three out of five).
In the following sections, we use \textit{they/them} to refer to the participants for anonymity.

\paragraph{Procedure.} 
The formative study consists of three sessions, lasting an hour in total for each participant. 
The first session gathered demographic information and provided background and the study's purpose. 
The second session includes three question-answering tasks focused on different topics.
The topics and questions were selected and formed from the Wiki-Bio dataset~\citep{manakul2023selfcheckgpt}, like a famous person's biography or an organization's history, that showcases GPT-3's varying levels of accuracy in generating information. 
To obtain the answers, the participants must refer to the generated text based on a given prompt, ``\textit{This is a Wikipedia passage about \{topic\},}'' using OpenAI Playground. 
They were informed that the model might generate inaccurate information and were encouraged to validate the answers in multiple generations. 
Afterwards, they were asked to indicate their level of confidence in their response. 
In the final session, we conducted a semi-structured interview with users to gain insight into the challenges they faced during the study and the reasons behind their justifications. 
Additionally, we asked about their desired features for a more efficient system. 
The formative study was conducted over a videoconferencing platform, where we observed the participants' behaviors of interacting with the OpenAI Playground from their screen sharing.
After the study, each participant was awarded \$25.

\subsection{Findings}
Following the observed behaviors and verbal feedback during the interviews, we summarize the findings as follows.

\paragraph{Token probability is not enough} 
Four participants, including all those with NLP expertise, examined the token-level log probability to determine the validity of the generated content (see Example \ref{example:token-level-logprob}). 
They all agreed that the model's confidence in the semantics cannot be justified solely based on the log probability. 
P1 commented that ``\textit{a token with low log probability doesn't necessarily indicate a lack of confidence in the semantics. Sometimes, it could simply be a matter of wording, like footballer vs. football player.}''
P4 shares a similar opinion. 
They mentioned that they would not readily trust the generation, even if the log probability of certain tokens is high. 
``\textit{I will examine the context. It's only when all relevant tokens have a high probability that I will feel confident in the statement}'' (P4). 
After considering all viewpoints, we conclude that the straightforward and commonly utilized metric of log probability is insufficient for users to comprehend the level of confidence in the generated text from a semantic perspective. 
\vspace{-2mm}
\newcommand{\hlcA}[2][yellow]{{%
    \colorlet{tmp}{#1}%
    \sethlcolor{tmp}\hl{ \,\,#2\,\, }}%
    \hspace*{-1mm}
}
\newcommand{\hlcC}[2][yellow]{{%
    \colorlet{tmp}{#1}%
    \sethlcolor{tmp}\hl{ \,\,#2\,\, }}%
    \hspace*{-1.4mm}
}
\newcommand{\hlcB}[2][yellow]{\raisebox{0.2mm}{\scalebox{0.7}{{%
    \colorlet{tmp}{#1}%
    \sethlcolor{tmp}\hl{ \large #2 }}%
    \hspace{-0.5mm}
}}}

\begin{figure}[htbp]
\textexample[]{}
{Token-level log probabilities (\hlcB[RoyalBlue4!5]{low} and \hlcB[RoyalBlue4!40]{high}) are not always understandable and informative to users.}
{\label{example:token-level-logprob}

\it
\hlcA[RoyalBlue4!5]{Rodrigo} 
\hlcA[RoyalBlue4!35]{Hern\'andez}
\hlcA[RoyalBlue4!54]{is} 
\hlcA[RoyalBlue4!54]{a} 
\hlcA[RoyalBlue4!60]{famous} 
\hlcA[RoyalBlue4!35]{Spanish}
\hlcA[RoyalBlue4!25]{professional}
\hlcA[RoyalBlue4!45]{footballer}
\hlcA[RoyalBlue4!20]{who}
\hlcA[RoyalBlue4!20]{currently}
\hlcA[RoyalBlue4!50]{plays}
\hlcA[RoyalBlue4!53]{the}
\hlcC[RoyalBlue4!50]{central}
\hlcC[RoyalBlue4!15]{midfielder}
\hlcC[RoyalBlue4!35]{for}
\hlcC[RoyalBlue4!15]{Premier}
\hlcC[RoyalBlue4!35]{League}
\hlcC[RoyalBlue4!50]{club}
\hlcC[RoyalBlue4!20]{Manchester}
\hlcC[RoyalBlue4!52]{City}
\hlcC[RoyalBlue4!50]{and}
\hlcC[RoyalBlue4!54]{the}
\hlcC[RoyalBlue4!40]{national}
\hlcC[RoyalBlue4!53]{team}
\hlcC[RoyalBlue4!50]{.}
}
\end{figure}
\vspace{-3mm}

\paragraph{Comparing multiple samples is challenging} 
We observed that all participants generated up to five samples per question. 
P4 suggested that if the initial three samples agree with each other, they intend to trust the model. They are aware that three samples may not be enough to infer confidence.
However, comparing more than three samples is beyond their capability. 
``\textit{Even if given ten samples, I would still only select three at random for inspection}'' (P4).
P2 also indicated that it is not easy to compare multiple samples. ``\textit{I have to check history results back and forth, memorize everything, which is time-consuming}'' (P2).
To enhance the situation, P5 suggested that we implicitly annotate the text's differences and provide a high-level summary.

\paragraph{Context helps users justify the reliability} 
Based on our observation, the majority of participants attempted to assess the overall credibility of the information generated before verifying each individual claim.
P1 said, ``\textit{If the model is not confident with key information, like a person's profession, I will not trust the whole generation anymore.}'' 
P2 expressed a similar opinion.
They suggested that before proceeding, they should assess the overall credibility of the information provided by comparing the first few generations. 
We also observed that P4 chose to ``Least Likely'' option to highlight the tokens with low Log probability in the first generation.
``\textit{If I saw a considerable proportion of tokens being highlighted, I knew that the model was making up facts.}''
For concrete information, we found that the participants not only verified the corresponding sentences but also read the context and check related claims. ``\textit{If the model is confident about the movie names, I would be more sure that the model makes her profession correct.}'' (P3).

\paragraph{Users have different levels of “tolerance” with inconsistency} 
After observing the inconsistency between the samples, the participants made varying decisions on whether to trust the generated information. 
P4 and P5 chose to trust the model to some extent, as long as they found more supporting evidence than contradictory ones. 
P1, however, rejected the claims that only occasionally occur in the samples, even though they did not find contradicting evidence. ``\textit{If the information does not occur frequently, either it is not that relevant or the model is less confident about it}'' (P1). 
Unlike the previous two participants, P2 has a medium-level ``tolerance''. 
They suggested that any claims with contradicting evidence should be rejected, but they could tolerate the absence of claims in some samples.

\subsection{Design Requirements}

We propose the following design requirements inspired by the findings from the formative study and related literature. 

\smallskip

\begin{enumerate}[leftmargin=0em, itemindent=1.7em, nolistsep, label=\textbf{R\arabic*}]
\setlength{\itemsep}{10pt}

\item \label{req:semantic-conf}
\textbf{Measuring confidence at the semantic level.}
The perplexity-based metrics, like token-level log probability, do not always imply the model's confidence in the generated information because there can be semantic equivalence between different sentences. 
Highlighting the tokens with the highest or the lowest probability is not very helpful \citep{kuhn2022semantic} and findings from our formative study. 
Rather than evaluating each individual word, it would be more effective to assess and convey the model's level of confidence in each statement from the semantic perspective, which requires a careful design of computational algorithms and visual representations.

\item \label{req:context} 
\textbf{Annotating the confidence in text for in-context understanding.}
Our findings show that the context helps users verify the reliability of specific claims. Rather than just presenting a list of claims, our goal is to provide context by including confidence information within the original text. This annotated text not only gives users a better understanding of the overall quality but also allows them to identify potential misinformation more easily.

\item \label{req:interaction}
\textbf{Flexible interactions to select claims.}
The formative study shows that users have different strategies and standards for searching and verifying inaccurate claims. As an illustration, certain users expect to find confident claims to assess the general quality of the output, while others may concentrate more on the suspicious claims that are rarely mentioned in other examples at the start.
There may not be a universal schema in highlighting to help users find their claims of interest. 
The system should be customizable, such as allowing users to select and highlight claims of interest. 

\item \label{req:evidence}
\textbf{Linking the confidence scores with evidence.} 
We observed that no user justified the reliability of generated text using token scores (token log probability in our study) alone.
Even though the scores are high, users may still generate multiple samples to find the evidence~\citep{cheng2021vbridge}. 
This motivates us to interactively link semantic confidence scores with evidence (i.e., supporting claims and contradicted claims) from other samples. This enables users to easily query and manage evidence for claims.

\end{enumerate}

\section{Methodology Overview}\label{sec:methodology}
To tackle the above challenges and fulfill the design requirements, we propose a mixed-initiative approach by incorporating a novel self-consistency-checking pipeline with a tailored visual interface.

Our formative study suggested that token-wise log probability as a model confidence measure is hard to interpret for end users.
In previous studies~\cite{kuhn2022semantic}, this confusion is attributed to the linguistic phenomena of semantic equivalence---sentences with different tokens and syntax structures can still convey the same meaning (e.g., \textit{player of football} and \textit{soccer player}).
To fulfill \ref{req:semantic-conf}, we follow the concept of \textit{Semantic Uncertainty} introduced by \citet{kuhn2022semantic}, which treats the generated text as a whole and measures the semantic variations between the possible generations.
Technically, it is achieved by sampling multiple LLM responses using the same prompt, where the samples are independent from each other. 
We then evaluate the consistency between these samples by assessing their logical entailments. 
The proportion of the supporting samples (i.e., samples that entail the focal response) indicates the LLM's
level of confidence.
A generation lacking confidence may be inaccurate and should not be trusted.
Using this idea, we propose a computational pipeline in Section ~\ref{sec:algorithm}, where we make two important extensions to make the algorithm scalable to long-form text and provide fine-grained explanations to the confidence judgments (\ref{req:evidence}). 

To facilitate user access, understanding, and interactive inspection of the self-consistency, we propose a visual interface, introduced in Section ~\ref{sec:interface}.
The user interface presents various levels of abstraction for the self-consistency information, where users can interactively drill down to specific claims (\ref{req:interaction}), inspect their alternatives, and find supporting evidential sentences in other sampled responses (\ref{req:context}). 

\begin{figure*}[htbp]
\centering
\includegraphics[width=0.9\linewidth]{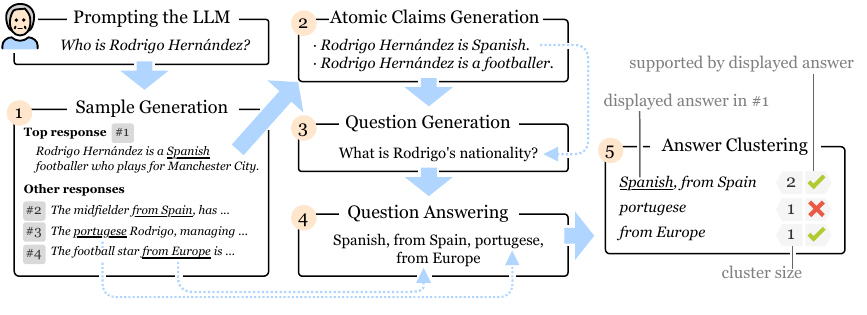}
\captionof{figure}{Pipeline of computing the self-consistency of individual claims (Algorithm \ref{algo:algo_main}). The input is a user prompt with which an LLM is invoked multiple times.
Afterwards, the atomic claims are generated based on the top text response, then turned into questions and answered by all other generations. The answers are then clustered together based on their meaning (e.g., \textit{Spanish} and \textit{from Spain}).} %
\label{fig:algo_main}
\Description{Figure 2 shows a computational pipeline we proposed in this paper that computes the self-consistency of individual claims. The input is a user prompt with which an LLM is invoked multiple times. Afterward, the atomic claims are generated based on the top text response, then turned into questions and answered by all other generations. The answers are then clustered together based on their meaning (e.g., Spanish and from Spain).}
\end{figure*}

\section{Self-Consistency-Checking Algorithm Design}
\label{sec:algorithm}

We propose a computational pipeline (Figure \ref{fig:algo_main}) that measures the semantic-level confidence (\ref{req:semantic-conf}) and links the confidence judgments with evidence (Example \ref{req:evidence}) to support users in assessing the confidence of natural language generations (NLGs).
The full algorithm is presented in Algorithm \ref{algo:algo_main}, and this section will explain and justify its steps.
A quantitative analysis of this algorithm as a tool for hallucination detection is presented in Appendix \ref{sec:experiment}.

\subsection{Assessing Semantic-level Confidence}
\label{sec:alg-semantic-confidence}

As introduced in Section \ref{sec:algorithm}, we follow the concept of S\textit{emantic Uncertainty} introduced by \citet{kuhn2022semantic}, where we treat the generated text as a whole and measure its confidence level in the semantic space (\ref{req:semantic-conf}).
In our initial approach, we generate multiple samples $S$ using the language model with the same prompt. 
We use an off-the-shelf natural language inference model to evaluate whether they entail the focal text generation (usually the top generation $S_1$). 
The proportion of the supporting samples (i.e., samples that entail the focal generation) indicates the level of confidence.

\begin{figure}[htbp]
\textexample[]{\linewidth}
{Breakdown of a long-form generation into atomic claims. 
The sentence can be correct in all, none, or some of them, which is hard to be represented with a single score.
}
{\label{example:atomic-claims}
\textbf{Generation}:
\textit{Rodrigo Hernández is a Spanish professional footballer who currently plays as a central midfielder for Premier League club Manchester City.}\\
\textbf{Atomic Claims}: \\
\it
\vspace{-2mm}
\begin{itemize}[left=0mm]
\item Rodrigo Hernández is Spanish.
\item Rodrigo Hernández is a professional footballer.
\item Rodrigo Hernández plays as a central midfielder
\item Rodrigo Hernández plays for Manchester City.
\item Rodrigo Hernández plays for a Premier League club.
\end{itemize}
}
\end{figure}

\paragraph{Scaling to Long-form Text.}
\label{sec:claim-gen}
The major limitation of the above method is the lack of scalability with long-form NLGs. 
When the generated text gets more complex, it is less likely that the two generations will convey exactly the same meaning.
In this case, we can expect that additional samples will either contradict or be neural to the focal generation. 
Therefore, for long-form text, it is not sufficient to use a single state to describe their logical relationships. 
Instead, our goal is to provide detailed explanations of which parts are confident and which are not by separating the semantics.
We break the text into atomic claims~\cite{min2023factscore}.
Example \ref{example:atomic-claims} illustrates how even a seemingly uncomplicated sentence can be composed of many atomic claims.
It is unlikely that all samples contain the exact same set of atomic claims.
We use InstructGPT to generate (Line 3 of Algorithm \ref{algo:algo_main}) atomic claims from each generated sentence with prompting.

\subsection{Linking Judgements with Evidence}
\label{sec:alg-question}

The confidence score alone may not always be intuitive or understandable to users. 
Additionally, we should enable users to easily access sentences from other samples that contain supporting or contradicting information (\ref{req:evidence}). 
Especially when an atomic claim lacks support from most samples, we should suggest alternative possibilities, i.e., what else can be true. 
For instance, if the majority of the samples do not support the claim ``\textit{Rodrigo Hernández Cascante is Spanish},'' what other nationalities could Rodrigo have? How many samples support each option?
To tackle the above complex task of natural language understanding, researchers typically create a schema for structured information extraction and employ semantic parsing algorithms to locate answers within documents.
We propose an innovative method of utilizing prompts to convert atomic claims into natural language questions, which are then used to query answers from other samples for comparison.

\begin{figure*}[ht]
\begin{minipage}{\linewidth}
\begin{algorithm}[H]
\includegraphics[]{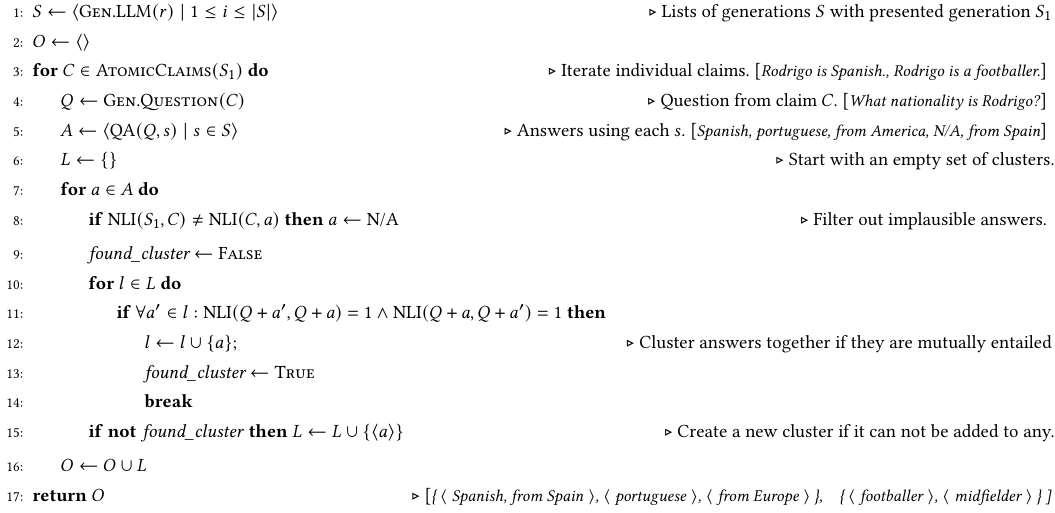}
\caption{%
    \textbf{Input}: Prompt $r$, number of generations $|S|$.\newline
    \textbf{Output}: List (for each claim) of sets (for each span) of clusters of semantically equivalent answers.
}
\label{algo:algo_main}
\end{algorithm}
\end{minipage}
\begin{center}
\vspace{2mm}
{\bf \raggedright Algorithm 1: The self-consistency checking algorithm (visualized in Algorithm \ref{fig:algo_main}) takes a user prompt, e.g., \textit{Tell me about Rodrigo Hernández.} as the inputs and outputs clusters of atomic claims in the responses.\par}
\end{center}
\vspace{3mm}
\end{figure*}

\paragraph{Question generation}
\label{sec:alg-question-gen}
We prompt the InstructGPT model to generate (such as What and When) natural language questions from each claim, e.g., ``\textit{What is Rodrigo's nationality}'' (Line 4 of Algorithm \ref{algo:algo_main}). 
The questions are then used to retrieve all plausible answers regarding each piece of information (i.e., atomic claims) in the presented response.
To ensure the questions are clear and relevant to the claim, we validate their answerability using an off-the-shelf question-answering (QA) model (Line 5). 
The answer is accepted only if the focal sample entailment of the claim is the same as the claim's entailment of the answer  (Line 8).

\paragraph{Answer querying}
We utilize the aforementioned QA model to obtain multiple answers from different samples (Line 5 of Algorithm \ref{algo:algo_main}). 
The QA model outputs answer spans, allowing us to easily identify the sentence containing the relevant information. 
Note that the QA model may not always be able to provide the answer for every sample, which we take into account when presenting to the user.
In addition to the limitations of the QA model, it is possible that the sample may not contain or indirectly refer to the required information. 
In these situations, users need to check these samples to find the evidence. 

\vspace{-2mm}

\begin{figure}[htbp]
\textexample[]{\linewidth}
{Support, contradiction, neutral, and equal relationships between answers.}
{\label{example:entailment}
\it
\textbf{\#1}: What is Rodrigo's nationality? Spanish\\
\textbf{\#2}: What is Rodrigo's nationality? European\\
\textbf{\#3}: What is Rodrigo's nationality? Portuguese\\
\textbf{\#4}: What is Rodrigo's nationality? from Europe\\[-0.7em]

\normalfont
\#1 supports \#2 but contradicts \#3 \\
\#2 is neutral to \#1 but equals \#4

\vspace{0.5mm}
}
\end{figure}

\vspace{-2mm}

\paragraph{Compare Answers}
In the last step, we cluster the answers with the same meaning. 
We define the semantic equivalence between two answers as whether the two answers support each other under the context of the question.
Practically,
we concatenate the question with each unique answer (see Example \ref{example:entailment}) and measure the entailment relationships between each pair of answers using the natural language inference (NLI) model introduced in Section \ref{sec:alg-semantic-confidence}.
We then group the semantically equivalent answers into clusters (Line 10 of Algorithm \ref{algo:algo_main}) and count the number of samples supporting each cluster.
Each answer group will be further classified into four types: \textit{equal}, \textit{support}, \textit{contradiction}, or \textit{neutral}, based on the corresponding atomic claim (see Example \ref{example:entailment}).

\section{The \system System}
\label{sec:interface}

We propose \system, an interactive system that helps LLM users identify and steer the inaccurate information in the generated text by inspecting the variations between multiple responses. 
\system consists of the \view{Prompt Inputter} (\autoref{fig:teaser}A), the \view{Response View} (\autoref{fig:teaser}B), the \view{Claim View} (\autoref{fig:teaser}C), and the Evidence View (\autoref{fig:teaser}D). 

The \system workflow begins with entering the prompt and obtaining the model's responses. 
Then, the users review the top response in the \view{Response View} and examine the \view{Keyword Annotations} (e.g., \autoref{fig:teaser}-1) to form an overall impression of the quality of the generated content.
In case the users conclude that the quality of the generation is unsatisfactory, they will reject the entire content.
Otherwise, the users go through the detailed information, inspect individual claims (in the \view{Claim View}), and check their supporting and contradicting evidence (in the \view{Evidence View}). 
When finding unsupported information, users use the editing function to remove the untrustful claims and get new annotations. 
The verification and correction process ends when users feel confident about the remaining information.

\subsection{Response View}

The \view{Response View} is the key component of \system, which displays the model response regarding the given prompt. 
We explored the design space of confidence visual annotation (\ref{req:context}) and proposed three visualization and interaction designs, including \view{Keyword Annotation}, \view{Brushing for Questioning}, and the \view{Editing} function.

\subsubsection{Keyword Annotation}
~\label{sec:keyword}
As introduced in Section \ref{sec:alg-question}, we retrieve the answers from the top response and additional samples regarding the natural language questions generated from atomic claims. 
The answers from the presented response, typically nouns or noun phrases, e.g., ``football player'', are considered to be keywords that convey concrete and important information.
The answers from the additional samples are considered alternatives, which are different options regarding the question. 

\begin{figure*}[htbp]
    \begin{minipage}{0.64\linewidth}
    \centering
    \includegraphics[width=0.85\linewidth]{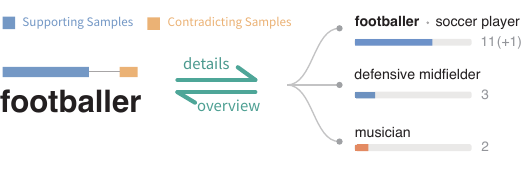}
    \captionof{figure}{The \view{Keyword Annotation} uses a small word-scale visualization (left) to display the proportions of different categories of samples or alternatives. Upon clicking, users are able to view a list of alternatives and inspect their details (right).}
    \label{fig:keyword}
    \Description{Figure 3 shows the design for the Keyword Annotation. We use a small word-scale visualization to display the proportions of different categories of samples or alternatives. Upon clicking, users are able to view a list of alternatives and inspect the detailed options.}
    \end{minipage}
    \hfill
    \begin{minipage}{0.33\linewidth}
        \centering
        \includegraphics[width=0.9\linewidth]{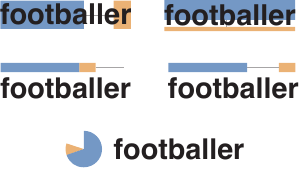}
        \captionof{figure}{The alternative designs for the \view{Keyword Annotation}. The final design is on the right side of the second row.}
        \label{fig:alternative}
        \Description{Figure 4 shows the alternative designs for the Keyword Annotation.}
    \end{minipage}
\end{figure*}

To convey the divergence between keywords and their alternatives, we propose a small word-scale visualization design that utilizes bars to display the proportions of different categories of alternatives (\autoref{fig:keyword} left).  
The length of the bar encodes the proportions of samples, providing \textcolor{tblue}{supporting}, \textcolor{torange}{contradicted}, and \textcolor{gray}{neutral} alternatives. 
The colors used are designed to be friendly for people with color blindness.
The empty space in the axis represents the samples where such information does not occur and is implicitly mentioned so that no corresponding tokens can be found. 

After clicking a \view{Keyword Annotation}, users can inspect the details of its alternatives through an options list embedded in the text (\autoref{fig:keyword} right).
Each option in the list represents a cluster of alternatives with the same meanings. 
A bar encodes the number of corresponding samples for each option, and its color indicates the entailment relationship between the alternatives and the primary answer. 
Some alternatives support the focal answer but do not semantically equal the focal answers. 
These alternatives may carry additional information (e.g., defensive midfielder vs. footballer) to which users may pay more attention. 

\paragraph{Rationale}
When designing the \view{Keyword Annotation}, our main consideration is to guarantee its visual efficiency and intuitiveness and minimize its effects on the original text layout. 
We considered four additional design alternatives. 
The initial two designs that overlaid bars on the keywords (\autoref{fig:alternative} row-1) were rejected due to the need to distort the visualizations to fit the length of the text.
As a result, longer keywords appear more salient than shorter ones, giving inaccurate visual hints to users.
We also considered juxtaposing a pie-chart-based glyph to indicate the proportions (\autoref{fig:alternative} row-3). However, this design is less efficient than using bars for encoding, especially when small slices exist. 
Finally, we compared two designs that pose bars above the annotated text (\autoref{fig:alternative} row-2), similar to Wikipedia's link text annotations~\citep{goffin2014exploring}. 
We chose the current one since it helps users better distinguish the number of supported samples (indicated by the blue bar) and the number of unsupported samples (indicated by the orange bars).

\subsubsection{Brushing for Questioning}
\label{sec:design-brush}
The data for the \view{Keyword Annotation} are generated based on an automatic computation pipeline introduced in Section~\ref{sec:alg-question}. 
However, it's important to notice that the algorithms used are not and may never be perfect in identifying all keywords that contain verifiable information. 
To supplement this, we propose an interaction named \view{Brushing for Questioning} to help users gain more power in controlling the computational pipeline.  
When the marking mode is activated, users can brush the text segment of interest (\autoref{fig:teaser}-2). 
According to the selected information, the system will generate a corresponding question to describe the text segment, which is then used to search for related information in other samples. 
Once users confirm the question, the system will use it to search for answers in all the samples and create a new \view{Keyword Annotation} based on the results (\autoref{fig:teaser}-3). 

\subsubsection{Editing the Generation}
The system allows users to edit the generated text and gain new verifications. This interaction allows users to conduct what-if analysis to test different editing strategies and view the changes on self-consistency. 
From iterative interactions with the system, the users can gradually validate and correct each piece of information in the generated text. 

\subsection{Claim View}
The \view{Claim View} displays the atomic claims derived from the generated text. We use a straightforward encoding to present the claim-related information (\autoref{fig:teaser}C).
For each individual claim, we presented the number of supporting, contradicting, and neutral samples. The color encoding is as same as the \view{Response View}. 
Long text generation contains numerous atomic claims (see the statistics in Appendix~\ref{sec:exp-res}). To help users easily locate claims of interest, we link the \view{Claim View} and the \view{Response View} through interactions (\ref{req:interaction}). Users can click on a sentence to filter the claims that are all derived from the focal sentence.

\subsection{Linking with Evidence}
The \view{Evidence View} (\autoref{fig:teaser}D) displays the additional samples, which is interactively linked to other views for selecting and highlighting evidence sentences and words (\ref{req:evidence}).
By default, the \view{Evidence View} displays all additional samples. 
When users select \view{Keyword Options} and \view{Claims}, the \view{Evidence View} will select and display only the relevant samples, where the related sentences and words are highlighted.
For example, when users select the ``footballer'' option, the \view{Evidence View} will only show samples that implicitly contain ``footballer'' or semantically equivalent words. 
Additionally, the system highlights the sentence and words that directly support and contradict the focal sentence or keyword. The color usage is consistent with the \view{Response View}. 
\subsection{System Implementation}
\label{sec:implementation}
\system{} comprises a user interface, a back-end server for data storage and processing, and two off-the-shelf language models deployed independently using an open-source framework~\cite{yao2023framework}. 
The interface is implemented through \textit{React} and \textit{D3.js}~\cite{bostock2011d3}. 
The back-end server is built upon \textit{Flask}.
We use \textit{spaCy} \citep{spacy2} for text tokenization and sentence segmentation.

The \system's computational pipeline (introduced in Section \ref{sec:algorithm}) is built upon three standard NLP tasks, including natural language inference (NLI), question generation, question answering (QA), and a novel task---atomic claim extraction. 
We use a \href{https://huggingface.co/microsoft/deberta-large-mnli}{\textit{DeBERTa}} \citep{he2021deberta} model fine-tuned with NLI tasks for justifying the logical entailment among text (Section \ref{sec:alg-semantic-confidence}) 
and a distilled \href{https://huggingface.co/deepset/tinyroberta-squad2}{\textit{RoBERTa}} model \citep{roberta} for the QA task (Section \ref{sec:alg-question}).
All models are adapted from the open-source versions in \textit{HuggingFace} with default hyperparameter settings.
To conduct atomic claim generation (Section \ref{sec:claim-gen}), we query OpenAI's \textit{text-davinci-003} model for completion (temperature of $0.7$) using a prompt simplified from Min et al.'s study ~\cite{min2023factscore}.
The question generation task (Section \ref{sec:alg-question}) is also accomplished using the same model and parameter settings except for the prompt.
All the prompts used in this study are listed in \autoref{sec:appendix-prompt}.

\section{Case Study}
\label{sec:case-study}

We evaluate the usability and usefulness of the system with ten LLM users.
The study design and result analyses are introduced in Section \ref{sec:user-study}. 
In this section, we present a case study based on the experience of one participant. 
We narrate the use case under a hypothetical scenario but with the users' real actions and feedback.

\paragraph{Hypothetical Scenario.} 
David (an alias) is an avid user of LLM. David was working on an essay about Funny Sculpture and was searching for relevant reports and stories. During his research, he came across a name he was unfamiliar with - Don Featherstone. Fueled by curiosity, David decided to learn more about Featherstone. He went to IntructGPT and typed the question, ``\textit{Tell me about Don Featherstone}'' (\autoref{fig:teaser}A).

\newcommand{\phrase}[1]{``\textit{#1}''}

\paragraph{Making an Overall Judgement}
After getting the response (\autoref{fig:teaser}B), David read through the passage. Based on this first impression of the model generation, he commented, ``\textit{The content seems coherent and reasonable.}''
Then he realized that the generated information may not be accurate, so he decided to thoroughly examine the details to determine the trustworthy content. He began his investigation by reviewing the annotated text (e.g., \autoref{fig:teaser}-1). 
Upon examination, he discovered that specific keywords, mainly from the former part of the passage, such as \phrase{2015}, \phrase{American}, and \phrase{Union Products}, were marked with blue bars, indicating that the model was highly certain about the accuracy of that information. Annotations for other keywords either have a significant orange bar or barely empty axes, indicating the lack of support from other samples. Then David decided to inspect the unsupported keywords one by one. 

\paragraph{Inspecting Evidences}
He first checked the birth year of Featherstone, which was \phrase{1936} according to the top response. ``\textit{It seems this information is controversial. Let's find out what other samples say.}''
After clicking the orange bar in the annotation (\autoref{fig:teaser}-1), David looked into the contradicting evidence sentences from other samples in the Evidence View (\autoref{fig:teaser}D). 
He noticed that the alternative statements about the person's birth year are diverse, ranging from 1920 to 1933.
David commented, ``\textit{If I have to choose, I will pick 1936 since over half of the samples support this number},'' and decided to trust the top response. 

\paragraph{Proposing a Question for the Information of Interest}
David then checked the rest of the sentence. He found that the sentence mentioned important information about the person's achievement (``\textit{best known for...}''), which was not automatically highlighted by the system. 
He remembered the Questioning by Brushing interaction and used this function to select the text segment of interest (``\textit{having created the iconic pink plastic flamingo lawn ornament}''). 
After two seconds, he saw the recommended question, ``\textit{What is Don Featherstone known for?}'' (\autoref{fig:teaser}-2) and confirmed it. 
After a few more seconds, he received alternative answers from other examples (\autoref{fig:teaser}-3). 
He noticed that the answers were all about Featherstone's creation---the pink plastic flamingo, even though they have different syntax structures.
David commented, ``\textit{I'm pretty certain about this statement},'' and confirmed this sentence. 

\paragraph{Verifying Associated Information.}
David proceeded to check the second sentence. 
In the sentence, he found that the information about Featherstone's birthplace and education experience are highlighted. 
``\textit{The birthplace and school are related information, [so I check them together]},'' David said. 
He clicked and expanded two keywords (\autoref{fig:teaser}-4 and \autoref{fig:teaser}-5). 
He first noticed the model somehow confidently predicted the person was born in \phrase{Leominster, Massachusetts}, but was uncertain about the person's school. 
After further inspection, David noticed that \phrase{Worcester} occurs both in an option of Featherstone's birthplace and a school name. 
He said, ``\textit{It strengthens the probability that Featherstone is born in Worcester [even though there is only one supportive sample]}.'' 
David decided to keep skeptical of the information mentioned in this sentence and commented, ``\textit{The only thing I'm sure is that the person was born in Massachusetts.}'' He simplified the sentence into ``\textit{Featherstone was born in Massachusetts.}''

\paragraph{Making Judgements according to the Context.}
David proceeded to review and edit the remaining generated content. He removed ``\textit{in 1956}'' in the third sentence due to the lack of support from other samples. For the following sentence (``\textit{After much trial and error...}'', \autoref{fig:teaser}-6), David noticed no keywords were highlighted. 
So he decided to investigate and clicked on the sentence, which brought up the Claim List (\autoref{fig:teaser}D). 
Upon review, David found that two of the three claims made in this sentence were well-supported by the majority of the samples (19 out of 20). 
The third claim was an extension of the second claim, which included the phrase "after much trial and error".  
``\textit{[The third claim] makes sense. I believe it is true [even though there are no supported samples]},'' David said and confirmed the sentence.
In the remaining time, David verified the last three sentences. He removed the phrase ``\textit{In 2007},'' which referred to the year Featherstone was inducted into the National Inventors Hall of Fame, and the last sentence due to a lack of support. 
After making the above edits, David was satisfied and confident with the remaining content.

\section{User Study}
\label{sec:user-study}

In this section, we present a user study to evaluate \system's usability and usefulness. We introduce insights about the participants' experiences of using \system in detecting and correcting misinformation from LLM outputs.

\subsection{Study Design}

\paragraph{Participants} In this study, we recruited ten participants (aged 25-54) with various backgrounds using an email list. 
All of the participants have used LLMs, with half of them using them almost daily. Additionally, all participants use LLMs through a visual interface, where the ChatGPT UI is the most popular one (used by nine out of ten participants).

\paragraph{Process} The study began with obtaining formal consent from the participants, followed by an introduction and demonstration session. 
During this session, we explained the study's goal and demonstrated the usage of the system. After this, the participants were required to perform a task that involved verifying the correct LLM generation of Don Featherstone's biography. 
They were encouraged to think aloud about their approach while performing the task. 
Once the task was completed, the participants were asked to fill out an exit survey to provide feedback on the system's usability and usefulness. 
We also conducted interviews with the participants to understand their overall experience using the system, including the reasons behind their actions, comparisons to conventional LLM interfaces, the usability limitations of the system, and their suggestions for improvements.
After the study, each participant was rewarded 25\$. 

\subsection{Result Analysis}

From our observations and the quantitative and qualitative feedback from users, we summarize the insights about the users' workflow on verifying and correcting the generated text, the comparisons with traditional LLM interfaces, the system's usability and usefulness, and the participants' desired improvements.

\subsubsection{Verifying and Correcting Generated Text.}
We observed participants' behaviors and comments in operating the system for fact verification and correction. During the interview, we further asked the participants to explain the reasons behind their actions. 
Combining the insights from the observation and qualitative feedback, we summarize the participants' common strategies and decision considerations as follows.

\paragraph{Verification} All participants agree that the number of supporting samples is the most essential visual hit for them to make the justification. However, not all participants interpret this information in the same way. 
The major difference lies in handling information that lacks support from most samples but is not contradicted by other samples. 
P3 commented, ``\textit{If the information is absent in most samples, it should be removed}.'' 
P10 suggested the same process as well. 
However, P6 told a different strategy, ``\textit{For concrete information, I'll keep it only if the score is high. For details, I can accept if there is no contradiction}.'' 
P5 suggested that their decision is based on the concrete scenario. 
If it is a very serious report, they will check all unsupported information. 
In general, all participants agree that categorizing the additional samples into \textit{support, neutral} and \textit{contradiction} helps them better understand the sample variations. 

Five participants also mentioned that they frequently use context information for deciding the truthfulness of certain information. 
P2 gave a concrete example that they thought the statement, ``\textit{Featherstone was born in 1936}'' (\autoref{fig:teaser}-1), is true because ``\textit{if you count his age [in the major events of his career], it totally makes sense.}'' 
P1 and P5 both suggested that the ``pink plastic flamingo'' occurs multiple times, so there is no way that Featherstone did not create it. 

\paragraph{Correction}
All participants edited the text more than once and compared the annotations before and after the edit. 
P10 suggested that for most unsupported information, they would try to improve it using alternative options. If it still doesn't work, they will confidently delete the information.
We also observed an unexpected correction strategy where the participants intentionally blurred the suspicious claim. 
For example, P3 found that the model is not confident when the person is introduced to the National Inventors Hall of Fame, but all possible answers derived from the samples are after 2000. 
So they changed the original text, 2007 to 2000s. 
Similarly, we observed two participants changed the birthplace from \textit{Leominster, Massachusetts} to \textit{Massachusetts} because all other potential answers are all cities in Massachusetts state (introduced in Section \ref{sec:case-study}). 

\subsubsection{Comparisons with Traditional LLM Interfaces}
During the interview, we asked the participants about the major differences between our system and the conventional LLM interfaces they have used regarding their user experience.

\paragraph{Alternative Responses Becoming More Accessible}
All participants expressed that when using ChatGPT UI or OpenAI Playground, they would never intentionally query the model multiple times and compare the responses. 
``\textit{It is mentally intensive to look back and force to compare these responses},'' commented P5. 
However, when using our system, they are more willing to inspect the sentences from other samples. 
P4 said, ``\textit{It is easy to navigate through other samples and see other possibilities.}''

\paragraph{Interacting with LLM Outputs in a Structured Way}
Four participants emphasized another significant contrast between the conventional LLM interfaces---our system allows users to interact with generated text, allowing them to conduct what-if analysis by editing the text. 
P6 commented, ``\textit{In OpenAI Playground, you cannot really 'interact' with the model}.'' 
They added, ``\textit{[Interactions through] text have too much freedom; I sometimes need interactions in a more structured way.}''
P10 emphasized that the editing functions allowed them to verify their ideas for corrections and gain more confidence to delete the information when all corrections didn't work. 

\subsubsection{Usability and Usefulness}

\begin{figure*}
\centering

\hspace{-13mm}
\includegraphics[height=4.6cm]{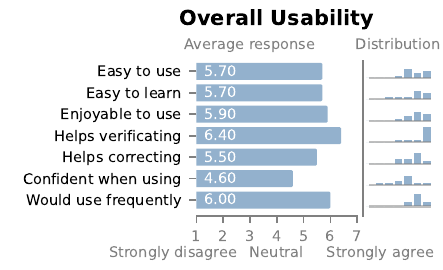}
\hspace{2.5mm}
\includegraphics[height=4.6cm]{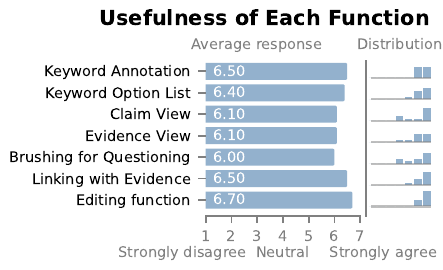}
\hspace{-10mm}

\vspace{-2mm}
\caption{Questionnaire results. Overall, the 10 participants reported high usefulness and satisfaction.}

\label{fig:questionaire}
\Description{Figure 5 shows the self-reported usability and usefulness of the ten participants. The left chart shows the overall usability scores: Easy to use (5.70 on average), Easy to learn (5.70), Enjoyable to use (5.90), Helps verification (6.40), Helps correction (5.50), Confidence when using (4.60), and Would frequently use (6.00). The right chart shows the usefulness of each function: Keyword Annotation (6.50), Keyword Option List (6.40), Claim View (6.10), Evidence View (6.10), Brushing for Questioning (6.00), Linking with Evidence (6.50), Editing function (6.70).}
\end{figure*}

This study measured the participants' self-reported usefulness and usability using a 7-point Likert Scale. 
We presented the results in \autoref{fig:questionaire}, showing that, overall, all participants found the system easy to use (5.70$\pm$1.00), easy to learn (5.70$\pm$1.27), and enjoyable to use (5.90$\pm$0.94). 
However, the participants expressed diverse opinions regarding confidence when using the system (4.60$\pm$1.36).
This can be partly explained by their lack of experience with a novel system, and we expect this confidence to rise during further usage.

\smallskip
Among all the components in the system, most participants (nine out of ten) suggested that they found the \view{Keyword Annotation} and \view{Alternatives} to be the most useful components. 
P3 commented that they had never seen something like this in language model outputs. 
``\textit{[The Keyword Alternatives] allows me to look into detailed options and figure out all possibilities},'' said P3. 
Despite Keyword Alternatives, P2 and P4 highly praised the \view{Linking to Evidence} interaction. 
P2 commented, ``\textit{The number [of supportive samples] only makes sense when I can check the corresponding sentences.}'' 
P5 suggested that their favorite interaction is editing the text and observing the changes in the \view{Keyword Annotation}. 
They said, ``\textit{I felt comfortable when seeing all orange bars turn into blue.}''
P10 suggested the systems allow them to inspect information with different levels for details and commented \textit{``The percentage [of supporting samples] of each fact is pretty visible. I know what to click if I want to go deeper.''}

\smallskip
Regarding the least useful component, four participants stated that they rarely used the \view{Claim View}. 
P2 mentioned that they only utilized the \view{Claim View} for the purpose of final checking. 
P4 initially forgot about this component but later remembered it after we gave him a hint. 
When asked about why they found the \view{Claim View} least useful, P1 suggested that it is easy to mentally create the atomic claims, and there is no need to get suggestions from model generation. 
While P4 said, ``\textit{The highlighted text takes almost all my attention. So, I ignored the claims.}''

The participants expressed diverse opinions toward the \view{Brushing for Question} interaction.
P1 favored this interaction and commented, ``\textit{It's fascinating how such complex interactions can be achieved through something as simple as brushing.}''
However, during our observation, we also noticed that not all participants found this interaction equally intuitive. 
We observed that four participants brushed the wrong text segment during the early exploration. 
For example, P4 attempted to figure out the year the pink plastic flamingo lawn ornament was created. 
He should have brushed the text ``1957'' but made a mistake by brushing the event text (``\textit{created the iconic pink...}''). 
As a result, they received undesired outputs. 
P6 suggested they would need more practice to get used to this interaction.

\subsubsection{Desired Improvements}  
\label{sec:desired_improvement}

We gathered the participants' feedback on the desired improvements as follows.

\paragraph{Encountering Evidence from External Resources.} 
Three participants (P1, P5, P6) mentioned that they expected the system to help them retrieve evidence not only from the additional samples but also from external resources. 
P5 expressed their skepticism regarding the reliability of the text generated by the model, emphasizing that even if the model consistently generated text, they would not fully trust it. 
``\textit{I have witnessed models confidently making errors},'' P5 stated, ``\textit{Therefore, I anticipated finding evidence from various sources to validate the generation}.'' 
On the other hand, P6 suggested using the system to cross-check the text generated by different language models on the same topic. 
P6 added that with this extension, the system can be used to help him assemble generations from multiple language models. 

\paragraph{Flexible Ways to Request Evidence.}
As discussed in the previous section, the \view{Brushing for Questioning} interaction is the most controversial design in our system. Three participants suggested that we had more flexible ways to propose questions and retrieve evidence. 
P4 suggested that not all questions should be necessarily tied to specific text in the generated content. 
Instead, P9 suggested that they expected to use a more straightforward interaction, like inputting the question in a separate window and searching for evidence. 

\paragraph{Intelligent Interactions for Text Editing.} 
P3 and P4 expressed their expectations of the system to provide intelligent interactions to automatically edit the text based on the suggested self-consistency. 
P3 mentioned that the system should be capable of making a trade-off between information enrichment and confidence. For instance, it should be able to provide a longer detailed answer, which might be wrong, or a shorter answer that is very confident. 
P4 and P7 also suggested a similar idea that the system should automatically select and present the most plausible information to users.

\section{Conclusion \& Discussion}
Our research addresses the intricate challenges of measuring when an LLM is producing nonfactual output and presenting this information to the users.
We achieve this by pivoting from token-level probabilities to claim-level confidences, which we estimate through self-consistency between multiple generations.
This allows us to present alternative keywords in the generation and we confirm the usability and usefulness of our system through a user study.

By using self-consistency as a measure of model confidence, our approach enables users to identify potentially inaccurate information.
This fosters a user-centered workflow for factual LLM text generation.
Our proposed pipeline is simple to use and potentially mitigates cases where users, soothed by the fluent and confidently presented LLM output, accept an untruthful answer.
In the rest of this section, we present discussions about the study's implications, applicability, limitations, and opportunities for future extensions.

\subsection{Design Considerations and Implications}
With the current surge of LLM usage and the integration of NLG outputs in different products, it is essential to understand the implications of this trend and balance a set of design considerations.  

\paragraph{Collaborative Human-LLM Interaction}
Human-AI interaction holds the potential for combining the strengths of both human and artificial intelligence in a collaborative setting~\citep{cheng2022polyphony,ming2019protosteer}. Specifically, the interaction with language models enables users to express themselves in plain language, with no need to learn a codified representation of specific knowledge or commands. Hence, designing expressive interactions and interfaces to facilitate this collaboration is essential.
Within that context, a crucial design consideration is to \textit{communicate that LLM outputs are uncertain, incomplete, or inconsistent}.
Utilizing mechanisms to integrate such information in the generated text directly or in additional cues around the text will facilitate a productive collaboration environment. 

\paragraph{The Illusion of LLM Understandability} 
Communicating in natural language has many obvious advantages; however, the immediate intelligibility of language can give a false sense of comprehension. Users could be lulled into a sense of understanding, thinking that the NLG process is based on similar procedures to human thought processes. 
This could lead to a harmful rationalization~\cite{sevastjanova2022beware} and a detrimental mental model about LLMs. 
In mitigating such effects, interfaces presenting LLM outputs also need to \textit{expose their hallucinations}. Similar to what we proposed in this paper, we need to highlight the possible disconnect of seemingly flawless language spawned by LLMs and the possibly flawed claims. 

\paragraph{Empowering Human Agency in NLG} 
In addition to explaining the LLM outputs and generation processes, another noteworthy design aspect is to consider the human capacity to control the outputs. Giving users the possibility to \textit{meaningfully provide feedback to the model} is vital to their perceived agency and, in turn,  to their contentment with the interaction interface. As pointed out by P6 in the previous section, having an interface that exposes the consistency of model decisions empowers the users to regain control of the text generation process. 

\subsection{Potential Usage Scenarios}
In the study, we use a hypothetical scenario of Wikipedia-style biography writing to motivate and evaluate the proposed system designs. 
Under this scenario, we present a case study to demonstrate how users can use \system{} to identify factual errors and make corrections.
Except for the above scenario, LLMs are also used to \textit{perform logical reasoning}, \textit{make predictions}, and \textit{suggest creative ideas}, where the generated text is not only about knowledge and facts but also includes predictions and creative content. 
The proposed consistency-based approach has the potential to be applied to complex scenarios to bring different insights.

\paragraph{Inspecting LLMs' Confidence in Predictive Tasks.}

When using LLMs to make predictions through open-ended questions, the content and the form of the generated responses can be diverse. 
Thus, it is hard to measure the model's confidence by token probability. 
A more general approach is to consider their semantic consistency and measure and communicate the model's confidence to users. 
This method can also be applied to the model's reasoning process to inform users of low-confident predictions and predictions with contradictory explanations, which should then be rejected.

\paragraph{Understanding the Divergence in Human-LLM Co-Creations.}

In human-LLM co-creations, such as story writing, users prompt the LLM to brainstorm creative ideas~\cite{ippolito2022creative}. 
In these scenarios, the proposed method has the potential to be extended to provide a visual summary of the diverse ideas from multiple LLM responses, showing the branches of the story using the proposed visualizations. 
The human-LLM co-creations are fundamentally different from the previous two scenarios, where users don't justify the correctness of the LLM response but want to know their diversities. 
So adapting the proposed system in such scenarios is challenging and requires explorations in future studies. 

\paragraph{Improving Awareness of Hallucinations in Human-LLM Conversations}

A major application of LLMs is seen in the development of general-purpose chatbots, like ChatGPT.
The proposed method can be extended to be used in conversational settings to identify hallucinations and present high-precision warning information to human users. 
This approach may increase users' awareness of the misinformation during the conversation and enable users to locate suspicious information for further verification.

\subsection{Limitations and Future Work}

\paragraph{Consistency does not always imply Truthfulness.}
Our approach is based on the assumption that high confidence in the generated text (represented by the consistency among responses) is a prerequisite for truthfulness, which is supported by previous studies~\cite{manakul2023selfcheckgpt,kuhn2022semantic} and our experiments presented in \autoref{sec:experiment}. 
However, from a different angle, high confidence is never a guarantee of truthfulness, as the model can be confidently wrong, e.g., due to out-of-date information. 
Besides, for long-form text responses, there is no perfect definition and measurement of consistency.
So the system may unintentionally produce misleading visual hints, e.g., showing that a claim is supported by very few responses, but it turns out to be true.
The users' interpretations of the presented confidence notion can be different, which may lead to unexpected results \citep{zouhar2021backtranslation}. %
To better understand how the presented consistency affects users' confidence and decisions, we plan to conduct user studies with real-world scenarios and decision-making tasks in future research.

\paragraph{Usage of Additional Language Models.}
In the proposed algorithm, we use off-the-shelf language models to analyze the semantic variations among the multiple responses. 
These language-model-enhanced methods will inevitably increase calculation expenses and may introduce errors and biases to the results. 
To minimize the cost and risk, we intentionally made the following approaches when developing \system.
First, we broke the complex algorithmic problem into elementary NLP tasks, including automatic claim extraction, entailment inference, question generation, and question answering, and applied widely-tested models or methods to perform these tasks as introduced in Section \ref{sec:implementation}.
Second, in the \system's interface, all the visual components based on computational results are interactively linked with the original instances (\ref{req:evidence}). 
Users can review and confirm the visual suggestions provided by the system.

Combining multiple language models or prompting methods to solve challenging tasks is becoming a common practice~\cite{wu2022ai,jiang2023graphologue}.
How to evaluate and justify their benefits and risks to human users are ongoing challenges concerning both AI and human-computer interaction fields, for which further studies are required.

\paragraph{Integrating Multiple Responses.}
Our approach aims to help users verify LLM responses, where we distinguish between the LLM's top response (i.e., the focal response) and other responses. 
We extract claims from the focal response and find supporting or contradictory evidence from other response samples.
Our approach does not consider the information presented in other responses but not represented in the focal response.
An interesting extension to develop a symmetric method that takes all responses' information into account and synthesizes an integration of all responses. 
To accomplish this, we need to address the challenges of breaking down responses (into atomic claims), verifying them, and logically and coherently reassembling the text.
We expect to conduct more explorations in future studies.

\paragraph{Verifying LLM Responses using Evidence from Additional Sources.}
From a technical perspective, the proposed algorithm can be extended to verify LLM responses using text from additional sources, such as a knowledge base, generations from other LLMs, and generations using paraphrased prompts, which is briefly discussed in Section \ref{sec:desired_improvement}.
To make use of external knowledge sources, the system should be integrated with information retrieval techniques, such as the computational pipeline in CRITIC \citep{gou2023critic}.
When a trustworthy external knowledge source does not exist, we could leverage the idea of Mixture of Experts (MoE) by verifying the focal generation with a set of models' responses from the same prompt.
Then, we can use the proposed algorithm to conduct the semantic variation analysis.
Similarly, we can use semantically similar prompts to make queries and compare LLM responses.
Retrieving evidence from additional sources poses its own visualization challenges, such as presenting knowledge provenances and out-of-context information, which requires further experiments and studies.

\begin{acks}
We acknowledge support from ETH Zurich, the Swiss National Science Foundation (Project No. 197155), and a Responsible AI grant by the Haslerstiftung.
\end{acks}

\bigskip

\bibliographystyle{misc/ACM-Reference-Format}
\bibliography{misc/reference_macros,bibliography}

\clearpage

\appendix

\section{Prompt Design}
\label{sec:appendix-prompt}

We use the following prompts to accomplish the atomic claim generation and question generation tasks.

\begin{figure}[htbp]
\textexample[]{\linewidth}
{The \textit{Atomic Claim Generation} Prompt (adapted from \cite{min2023factscore}) is used to generate atomic claims from each sentence of the presented response (Section \ref{sec:claim-gen}).}
{
\label{example:prompt-claim}
Please breakdown the following sentence into independent facts\\
Examples: \\

Input: In 1963, Collins became one of the third group of astronauts selected by NASA and he served as the back-up Command Module Pilot for the Gemini 7 mission.\\
Output:\\
- Collins became an astronaut.\\
- Collins became one of the third group of astronauts selected by NASA.\\
- Collins became one of the third group of astronauts selected by NASA in 1963.\\
- Collins served as the back-up Command Module Pilot.\\
- Collins served for the Gemini 7 mission.\\
\\
Input: In addition to his acting roles, Bateman has written and directed two short films and is currently in development on his feature debut.\\
Output:\\
- Bateman has acting roles. \\
- Bateman has written two short films. \\
- Bateman has directed two short films. \\
- Bateman has written and directed two short films. \\
- Bateman is currently in development on his feature debut. \\

Input: Michael Collins (born October 31, 1930) is a retired American astronaut and test pilot who was the Command Module Pilot for the Apollo 11 mission in 1969.\\
Output:\\
- Michael Collins was born on October 31, 1930.\\
- Michael Collins is retired.\\
- Michael Collins is an American.\\
- Michael Collins was an astronaut.\\
- Michael Collins was a test pilot.\\
- Michael Collins was the Command Module Pilot.\\
- Michael Collins was the Command Module Pilot for the Apollo 11 mission.\\
- Michael Collins was the Command Module Pilot for the Apollo 11 mission in 1969.\\

Input: During his professional career, McCoy played for the Broncos, the San Diego Chargers, the Minnesota Vikings, and the Jacksonville Jaguars.\\
Output:\\
- McCoy played for the Broncos during his professional career.\\
- McCoy played for the San Diego Chargers during his professional career.\\
- McCoy played for the Minnesota Vikings during his professional career.\\
- McCoy played for the Jacksonville Jaguars during his professional career.\\

Input: \texttt{\{the target sentence\}} \\
Output:\\
}
\end{figure}

\begin{figure}[htbp]
\textexample[]{\linewidth}
{The \textit{Question Generation} Prompt-I is used to generate questions from each claim (Section. \ref{sec:alg-question-gen}).}
{
Write a question from the sentence about the sentence subject.\\

context: He became an astronaut.\\
question: What did he become?\\

context: She became one of the third group of astronauts selected by NASA.\\
question: Who selected her to become one of the third group of astronauts?\\

context: He became one of the third group of astronauts selected by NASA in 1963.\\
question: When was he selected to become one of the third group of astronauts selected by NASA?\\

context: She was born on October 31, 1930.\\
question: When was she born?\\

context: \texttt{\{the target claim\}} \\
question:
}
\end{figure}

\begin{figure}[htbp]
\textexample[]{\linewidth}
{The \textit{Question Generation} Prompt-II is used to generate questions to support the \textit{Brushing for Questioning} interaction (Section. \ref{sec:design-brush}).
The output question describes the selected text from users.}
{
Write a question for the target words according to the context.\\

context: He became an astronaut.\\
target: astronaut\\
question: What did he become?\\

context: She became one of the third group of astronauts selected by NASA.\\
target: NASA\\
question: Who selected her to become one of the third group of astronauts?\\

context: He became one of the third group of astronauts selected by NASA in 1963.\\
target: 1963\\
question: When was he selected to become one of the third group of astronauts selected by NASA?\\

context: She was born on October 31, 1930.\\
target: October 31, 1930\\
question: When was she born?\\

context: \texttt{\{the target claim\}} \\
target: \texttt{\{the selected text\}} \\
question:
}
\end{figure}

\clearpage

\section{Quantitative Analysis of Self-Consistency}
\label{sec:experiment}

This section presents an algorithmic evaluation of how well the between-sample consistency can be utilized to identify inaccurate information in long-form NLGs. 
Additionally, this study examines the relationship between sample size and performance to determine the number of additional samples required.
Finally, we analyze the failure cases to investigate the limitations of consistency-based approaches in hallucination identification. 

\subsection{Experiment Design}

\paragraph{Dataset}
We use the dataset proposed by Min et al.~\citep{min2023factscore}, which includes passages about peoples' biographies generated from InstructGPT and ChatGPT. The passages are generated with the prompt ``\textit{Tell me a bio of \{entity\}},'' where the \textit{entity} refers to a person entity who has a corresponding Wikipedia page. 
In the dataset, each generation has already been divided into atomic claims, which were categorized as \textbf{S}upported (\textbf{\textcolor{tblue}{S}}), \textbf{N}ot \textbf{S}upported (\textbf{\textcolor{torange}{NS}}), or \textbf{Ir}relevant (\textbf{IR}) by human annotators.
We excluded unfeasible generations with no claims, e.g., ``\textit{I couldn't find any notable public figure named...}''. Among all generations, 181 InstructGPT and 157 ChatGPT generations were kept.
On average, each InstructGPT generation contains 26.1 claims (\textbf{\textcolor{tblue}{S}}: 44\%, \textbf{\textcolor{torange}{NS}}: 42\%, \textbf{IR}: 14\%), and each ChatGPT generation contains 34.6 claims (\textbf{\textcolor{tblue}{S}}: 59\%, \textbf{\textcolor{torange}{NS}}: 31\%, \textbf{IR}: 10\%). 
As per Min et al.'s study design~\citep{min2023factscore}, we eliminated all irrelevant claims. These claims are not connected to the factuality of the generation and cannot be verified with consistency between generations. 

\paragraph{Study Process}
We query additional samples ($n=20$) from the corresponding model for each generation with the same prompt. We then use the DeBERTa model to evaluate whether the sample supports the atomic claims or not. The Consistency Score of each claim is then calculated based on the proportion of samples that support the claim. We measure the performance by using the Consistency Score to predict the truthfulness of the claims (S or NS) in each instance. To evaluate the performance, we relied on the Area Under the Receiver Operating Characteristic (AUROC) metric. We chose this metric because there is no unique strategy to convert the Consistency Score into a binary prediction. Instead, AUROC measures performance across different thresholds.

To investigate the relationship between sample size and performance, we conducted additional experiments with sample sizes ranging from 1 to 20. The average results are reported in the following section.

\subsection{Quantitative Analysis}
\label{sec:exp-res}

\autoref{tab:exp-res} summarizes the results of the experiments. For the generations from the InstructGPT model, the average AUROC score is 0.856. In contrast, the generations from the ChatGPT model are more challenging to verify (AUROC=0.764) by only considering the consistency between multiple generation samples. 

\vspace{-3mm}
\begin{minipage}{0.95\linewidth}
\newcommand{\vartext}[1]{\,\,{\small ($\sigma^2{=}#1$)}}

\begin{table}[H]

\begin{tabular}{lcc}
\toprule
& \bf InstructGPT & \bf ChatGPT \\
\midrule
\bf AUROC & 0.856\vartext{0.134} & 0.764\vartext{0.137} \\
\bf \multirow{2}{*}{\shortstack[l]{Consistency}} \textcolor{torange}{NS} & 0.156\vartext{0.234} & 0.278\vartext{0.330} \\
\bf \hspace{18.5mm} \textcolor{tblue}{S} & 0.564\vartext{0.355} & 0.625\vartext{0.368} \\
\bottomrule
\end{tabular}
\vspace{1em}
\caption{The claims extracted from ChatGPT generations are more difficult to verify using the Consistency Score.}
\label{tab:exp-res}
\Description{Table 1 shows the AUROC and Consistency scores for the InstructGPT and ChatGPT generations. AUROC: InstructGPT (avg: 0.856, std: 0.134), ChatGPT (avg: 0.764, std: 0.137). Consistency Scores for Not Supported (NS) Claims: InstructGPT (avg: 0.156, std: 0.234), ChatGPT (avg: 0.278, std: 0.330). Consistency Scores for Supported (S) Claims: InstructGPT (avg: 0.564, std: 0.335), ChatGPT (avg: 0.625, std: 0.368).}
\end{table}
\end{minipage}
\hfill
\begin{figure}[H]
\includegraphics[width=\linewidth]{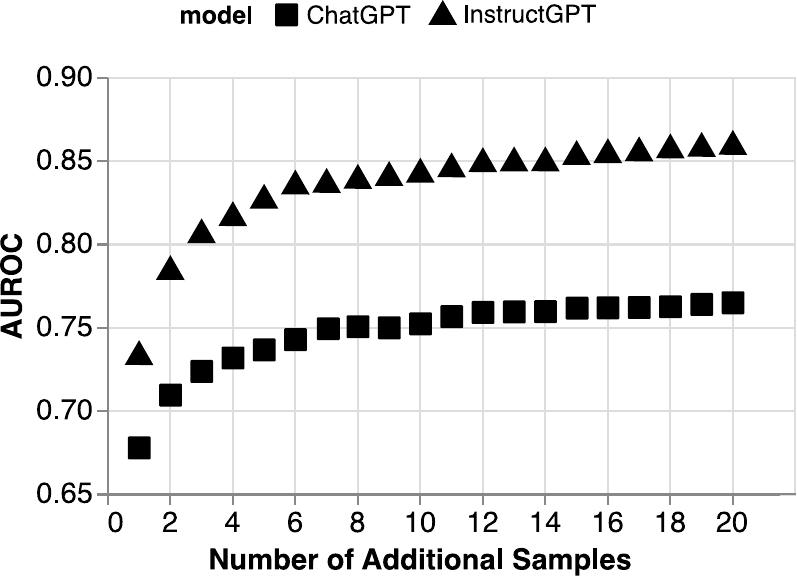}
\caption{Increasing the number of additional samples improves Consistency Scores, but with diminishing returns.}
\label{fig:exp-line}
\Description{Figure 6 shows the AUROC under different additional sample numbers (ranging from 1 to 20). For ChatGPT generations, the AUROC scores rise from 0.68 (n=1) to 0.76 (n=20). For the InstructGPT generations, the AUROC scores rise from 0.73 (n=1) to 0.86 (n=20). We found increasing the number of additional samples improves Consistency Scores but with diminishing returns.}
\end{figure}

\vspace{3mm}

\begin{figure*}[htbp]
\vspace{-2mm}
\centering
\includegraphics[width=\linewidth]{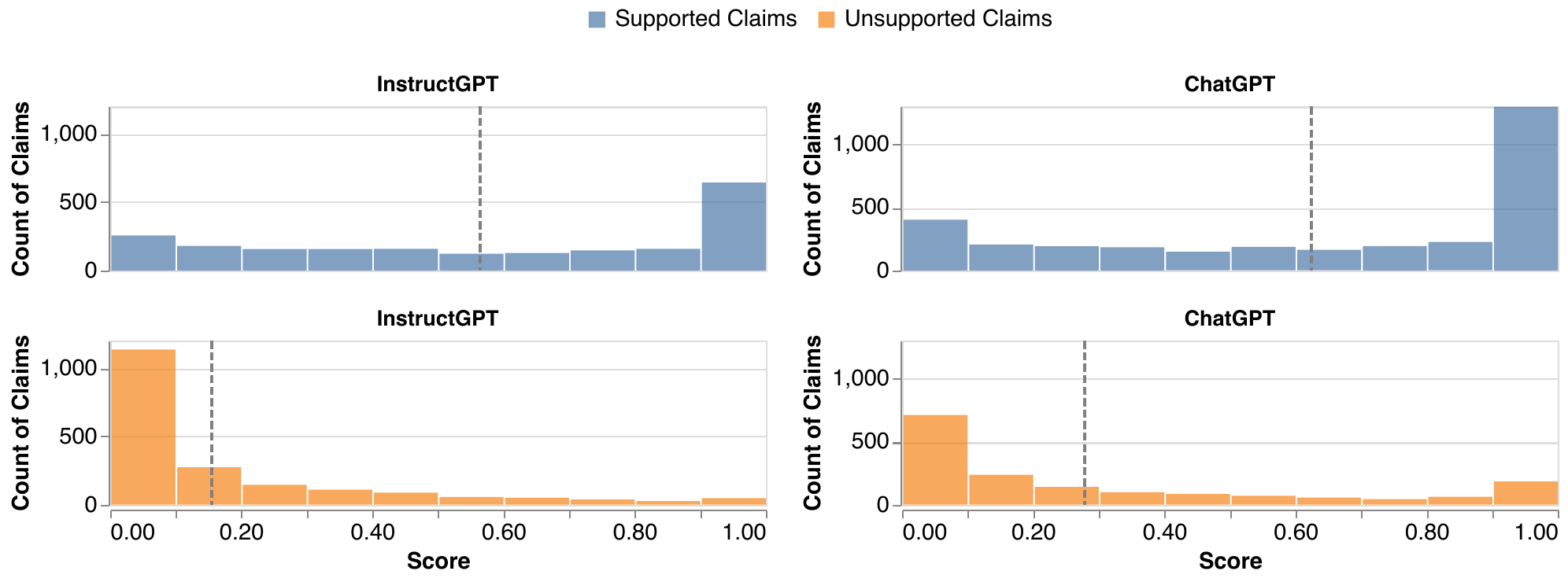}
\vspace{-3mm}
\caption{Based on the distributions of the Consistency Scores, it appears that most \textcolor{torange}{\textbf{NS claims}} (from the two models) have low scores. Conversely, \textcolor{tblue}{\textbf{S claims}} have a wider range of scores. 
The findings suggest that claims with high Consistency Scores (>0.5) are highly unlikely to contain inaccurate information. However, it's crucial for users to exercise caution when evaluating claims with low Consistency Scores, as they may still be true.} 
\label{fig:exp-hist}
\Description{Figure 7 shows the distributions of the Consistency Scores, it appears that most NS (not supported) claims (from the two models) have low scores. Conversely, S (supported) claims have a wider range of scores. The findings suggest that claims with high Consistency Scores (>0.5) are highly unlikely to contain inaccurate information. However, it’s crucial for users to exercise caution when evaluating claims with low Consistency Scores, as they may still be true.}
\vspace{-2mm}
\end{figure*}

We also present the average Consistency Scores (i.e., the proportions of the supportive samples) for the \textcolor{tblue}{\textbf{S claims}} (i.e., true claims) and \textcolor{torange}{\textbf{NS claims}} (false claims) in \autoref{tab:exp-res} and present the detailed score distribution in \autoref{fig:exp-hist}.
The results show that in generations from both two models, most \textcolor{torange}{\textbf{NS claims}} have low scores, indicating that most additional samples do not support these claims. Meanwhile, \textcolor{tblue}{\textbf{S claims}}' Consistency Scores have higher variance. 
The findings suggest that claims with high Consistency Scores (>0.5) are unlikely to contain inaccurate information.
However, users should still exercise caution when evaluating claims with low Consistency Scores, as they may still be true.

We evaluate the performance of the Consistency Scores with different numbers of additional samples (ranging from 1 to 20) and report the results in \autoref{fig:exp-line}. 
Consistency Scores perform better with more additional samples. 
For generations from the InstuctGPT model, three additional samples can guarantee a good performance (AUROC > 0.80).
As the number of additional samples increases, the marginal improvements become less significant.

\subsection{Qualitative Analysis}
We have shown \system helps users validate the atomic claims in NLG through benchmarks and user studies. \system utilizes the consistency across a sample of model generations as a measure of the model's confidence in a particular claim.
To better understand the limitations, we investigate incorrect claims for which the model consistency is high, i.e. cases where our algorithm mistakenly predicts that the claim is correct.

\paragraph{Protocol} 
First, we isolate the erroneous claims that achieved the top 10\% highest consistency among all erroneous claims, meaning that the model is confident about it.
We find 171 and 129 errors with the InstructGPT and ChatGPT models, respectively. 
We manually inspect each error, finding that 80 and 92 of the errors from the InstructGPT and ChatGPT models are mislabeled entries in the benchmark -- the claims are, in fact, true. 
Amongst the 91 and 37 true errors for the respective models, we annotate the key error categories. We structure the categories based on the person-entity \textit{types} (e.g. occupation, place of birth) that are annotated in Wikidata, the knowledge base that underlies Wikipedia. In some errors, the model entirely mistakes the true person-entity, and in others, the entity is correctly identified; however, specific details are mistaken. The breakdown of errors by category is summarized in Table \ref{tab:error_examples}.

\paragraph{Analysis} The \textit{Incorrect Person} category of errors is the most severe, wherein the generations should not be trusted at all. A frequent signal for this error bucket is that the model mis-states the person's profession.  We find 44\% (40) and 16\% (6) of the InstructGPT and ChatGPT errors, respectively, fall into this category. In the \textit{Incorrect Details} error categories, the model identifies the correct person but generates some incorrect atomic claims, such as the person's birthplace or place of education. The three key error buckets of incorrect details that we observe are generating (1)  numerical errors, (2) out-of-date facts, and (3) fuzzy claims. 

These errors align with the well-known limitations of current language models. Numeracy is challenging for popular language models since they use compression-based subword tokenization to represent numbers.
In these cases, numbers which occur frequently in the data, such as \texttt{1000000} are represented as a unique token, but a \texttt{7705453} might be represented as a sequence of \texttt{770$\cdot$}, \texttt{545$\cdot$}, and \texttt{3}.
This limits the model performance and \cite{wallace2022do} finds that Transformer-based models that use subword tokenization perform worse than character-based language models on tasks such as identifying the maximum number in text.
Next, retraining or updating large language models is increasingly expensive at massive parameter scales, which can result in the generation of out-of-date memorized facts over time \cite{zhang2021situated}. The fuzzy claims category includes statements that are mostly or fully true but vague and, therefore, difficult to validate. These vague claims are the least harmful to the user. In the few remaining errors, for instance, the person's birthplace or ethnicity, we observe the model occasionally predicts the person is from a relatively popular location over the true answer. For instance, an incorrect claim by InstructGPT is \textit{John Estes is from Nashville}, the Tennessee capital, whereas he was actually born in \textit{Ripley, Tennessee}. Prior work demonstrates a strong correlation between the number of times a fact is mentioned during pretraining and the model's ability to answer a fact-based question accurately \cite{kandpal2023larg}. 

Strategies such as retrieval-augmented generation \cite{lewis2021retrievalaugmented} may help us address aspects of these limitations in future work. We hope our analysis, showing that the currently popular language models are highly confident in these error buckets, can help users calibrate when to trust the generations and can guide future research on NLG verification.

\begin{table*}[htbp]
\begin{center}
    \vspace{3mm}

\resizebox{\linewidth}{!}{
\begin{tabular}{
p{1.7cm} p{4.9cm} p{6cm}
>{\hspace{-2mm}}c
>{\hspace{-2mm}}c
}
\toprule
 \bf Error  & \bf Examples & \bf Explanation & \bf \hspace{-2mm}InstructGPT & \bf ChatGPT\\
\midrule
Person & \emph{Zamfir Arbore} \newline \emph{Chaim Malinowitz is a speaker.} &  Model predicts mucisian, rather than the political activist. \newline Chaim Malinowitz is a rabbi, dayan, and scholar. & 44\% (40)  & 16\% (6) \\
\midrule
Details/ \newline  Numerical   &   \emph{Takeo Miki served as the 41st Prime Minister of Japan.}  \newline \emph{Chadwick Boseman was born on November 29, 1977.} &  Takeo Miki served as the 66th Prime Minister. \newline\newline Chadwick Boseman was born in 1976.  &  16\% (14)  & 19\% (7)  \\
\midrule
Details/ \newline Out of Date   &  \emph{Mauro Emanuel Icardi plays for Italian Serie A club Inter Milan.}  & Mauro Icardi plays for Süper Lig club Galatasaray. & 7\% (6) & 11\% (4)  \\
\midrule
Details/ \newline Fuzzy Claims & \emph{Song Kang has appeared in several films.} \newline \emph{John Estes is a multi instrumentalist.} &  For several, the claims may be true or partially true, but are vague or missing one detail. &  19\% (17) & 30\% (11) \\
\midrule
Details/ \newline Other   & \emph{Damir Memović was born in Bosnia and Herzegovina.} \newline \emph{Marie Alexandrine Becker was French.} &  Damir Memović was born in  Belgrade, Serbia. \newline Marie Alexandrine Becker was Belgian.  &  15\% (14) &  24\% (9) \\
\cmidrule{4-5}
&& \hfill \bf Total & 91 & 37 \\
\bottomrule
\end{tabular}
}
\end{center}
\caption{We identify the key error buckets for each model, when used in our system. We specifically focus on the examples with high-consistency across the model generations.}
\label{tab:error_examples}
\Description{Table 2 shows the key error buckets for each model. The detailed description of each key type can be found in A.3 Qualitative Analysis.}
\end{table*}

\clearpage

\end{document}